\documentclass[11pt]{amsart}
\usepackage{graphicx, amsmath, amssymb, latexsym, amsfonts, amsthm, amssymb,epstopdf,lscape}
\newtheorem{defin}{}
\newtheorem{saetze}[defin]{}
\newtheorem{conjec}[defin]{}
\newtheorem{lemmas}[defin]{}
\newtheorem{folger}[defin]{}
\newtheorem{bemerk}[defin]{}

\newenvironment{thm}  {\begin{saetze}\it {\bf Theorem:}}{\end{saetze}}

\begin{document}
\title[A CCA secure cryptosystem using matrices over group rings]
{A CCA secure cryptosystem using matrices over group rings}
\author[D. Kahrobaei]{Delaram Kahrobaei}
\address{CUNY Graduate Center and City Tech, City University of New York}%
\email{DKahrobaei@GC.Cuny.edu}
\thanks{Research of the first author was partially supported by a PSC-CUNY grant from
the CUNY research foundation, as well as the City Tech foundation.}
\author[C. Koupparis]{Charalambos Koupparis}
\address{CUNY Graduate Center, City University of New York}%
\email{ckoupparis@GC.Cuny.edu}
\author[V.Shpilrain]{Vladimir Shpilrain}
\address{The City College of New York and CUNY Graduate Center}
\email{shpil@groups.sci.ccny.cuny.edu}
\thanks{Research of the third author was partially supported by
the NSF grants DMS 0914778 and CNS 1117675}

\maketitle
\begin{abstract} We propose a cryptosystem based on matrices over group rings and  claim that it is
secure against  adaptive chosen ciphertext attack.
\end{abstract}

\section{Cramer-Shoup cryptosystem}

The Cramer-Shoup cryptosystem is a generalization of ElGamal's
protocol. It is provably  secure against adaptive chosen ciphertext
attack (CCA). Moreover, the proof of security relies only on a
standard intractability assumption, namely, the hardness of the
Diffie-Hellman decision problem in the underlying group (see
\cite{shoup}, \cite{shoup2}), and a hash function $H$ whose output
can be interpreted as a number in $\mathbb Z_q$ (where $q$ is a
large prime number). An additional requirement is that it should be
hard to find collisions in $H$. In fact, with a fairly minor
increase in cost and complexity, one can eliminate $H$ altogether.

\subsection{Definition of provable security against adaptive chosen ciphertext attack}
A formal definition of security against active attacks evolved in a
sequence of papers by Naor and Yung, Rackoff and Simon, Dolev, Dwork
and Naor. The notion is called {\it chosen ciphertext security} or,
equivalently, {\it non-malleability}. The intuitive thrust of this
definition is that even if an adversary can get arbitrary
ciphertexts of his choice decrypted, he still gets no partial
information about other encrypted messages. For more information see
\cite{shoup}, \cite{shoup2}.

We define the following game, which is played by the adversary.
First, we run the  enryption scheme's key generation algorithm, with
the necessary input parameters. (In particular, one can input a
binary string in $\{0,1\}^n$, which describes the group $G$ on which
the algorithm is based.) The adversary is then allowed to make
arbitrary queries to the decryption oracle, decrypting ciphertexts
which he has chosen.

The adversary then chooses two messages, $m_0$ and $m_1$, and
submits these to the encryption oracle. The encryption oracle
chooses a random bit $b\in\{0,1\}$ and encrypts $m_b$. The adversary
is then given the ciphertext, without knowledge of $b$.

Upon receipt of the ciphertext from the encryption oracle, the
adversary  is allowed to continue querying the decryption oracle. Of
course the adversary is not allowed to submit the output ciphertext
of the encryption oracle.

Finally, at the end of the game, the adversary must output
$b'\in\{0,1\}$,  which is the adversary's best guess as to the value
of $b$. Define the probability that $b'=b$ to be $1/2+\epsilon(n)$,
$\epsilon(n)$ is called the adversary's advantage, and $n\sim |G|$.

We say the cryptosystem is CCA-2 secure if the advantage of any
polynomial-time adversary is negligible. Note that a negligible
function is a function that grows slower than any inverse
polynomial, $n^{-c}$, for any particular constant $c$ and large
enough $n$.

\subsection{The Cramer-Shoup Scheme}\hfill \\
\noindent\textbf{Secret Key:} random $x_1, x_2, y_1, y_2, z \in \mathbb Z_q$\\
\textbf{Public Key:}
\begin{center}
group $G$; ~$g_1, g_2 \ne 1$ in $G$\\
$c = {g_1}^{x_1}{g_2}^{x_2}, ~d= {g_1}^{y_1}{g_2}^{y_2}$\\
$h={g_1}^z$.\\
\end{center}
\textbf{Encryption} of $m \in G$: ~$E(m)=(u_1, u_2, e, v)$, where
\begin{center}
$u_1 = {g_1}^r, u_2 = {g_2}^r, e = h^r m, v=c^r d^{r \alpha}$, where $r \in \mathbb Z_q$ is random,  and\\
$\alpha = H(u_1, u_2, e).$
\end{center}
\textbf{Decryption} of $(u_1, u_2, e, v)$:
\begin{center}
If $v= {u_1}^{x_1+ \alpha y_1} {u_2}^{x_2 + \alpha y_2}$, where $\alpha = H(u_1, u_2, e),$\\
then $m= e/{{u_1}^z}$\\
else ''reject"
\end{center}

\begin{thm} \cite{shoup} \label{shoup} The Cramer-Shoup cryptosystem is secure against adaptive chosen
ciphertext attack assuming that (1) the hash function $H$ is
 chosen from a universal one-way family, and (2) the Diffie-Hellman decision problem is hard in the group $G$.
\end{thm}

\section{A CCA-2 secure cryptosystem  using matrices over group rings}
\label{protocol}

In \cite{KKS}, the authors proposed a public key exchange using
matrices over group rings.  They offer a public key exchange
protocol in the spirit of Diffie-Hellman, but they use matrices over
a group ring of a (rather small) symmetric group as the platform and
discuss security of this scheme by addressing the Decision
Diffie-Hellman (DDH)  and
Computational Diffie-Hellman (CDH) problems for that platform.\\

Here we propose to use a similar platform and show that a  scheme
similar to the Cramer-Shoup  scheme is CCA-2 secure.  Our protocol
is as follows:

\noindent \textbf{Secret Key:} random $x_1, x_2, y_1, y_2, z \in \mathbb Z_n$\\
\textbf{Public Key:}
\begin{center}
$3\times3$ non-identity matrices $M_1, M_2 \in
M_{3\times3}(\mathbb{Z}_{7}[S_5])$  such that $M_1$
is invertible and $M_1 M_2 = M_2 M_1$\\
$c = {M_1}^{x_1}{M_2}^{x_2}, ~d= {M_1}^{y_1}{M_2}^{y_2}$\\
$h={M_1}^z$.\\
\end{center}
\textbf{Encryption} of a message $N \in
M_{3\times3}(\mathbb{Z}_{7}[S_5]) $: $E(N)=(u_1, u_2, e, v)$, where
\begin{center}
$u_1 = {M_1}^r, ~u_2 = {M_2}^r, e = h^r N, ~v=c^r d^{r \alpha}$, ~$r \in \mathbb Z_n$ is random, and\\
$\alpha = H(u_1, u_2, e).$
\end{center}
\textbf{Decryption} of $(u_1, u_2, e, v)$:
\begin{center}
If $v= {u_1}^{x_1+ \alpha y_1} {u_2}^{x_2 + \alpha y_2}$, where $\alpha = H(u_1, u_2, e),$\\
then $N=  ({u_1}^z)^{-1} e$ (Note that $u_1$ is invertible since $M_1$ is chosen to be invertible.)\\
else ''reject"
\end{center}

\noindent \textit{Remarks}: $M_1$ must always be chosen to be an
invertible matrix, whereas $M_2$ is  just any matrix such that
$M_1M_2 = M_2M_1$. One must also decide what group $ \mathbb{Z}_n$
to use, i.e., $n$ must be specified.

\section{Adaptive CCA security for matrices over group rings}

We aim to show, by using Theorem \ref{shoup}, that  if for
invertible matrices  over $M_{3\times3} \mathbb{Z}_{7}[S_5]$ the DDH
problem is hard, then the previously mentioned cyrptosystem is
secure against adaptive chosen ciphertext attack. More formally,

\begin{thm}\label{CCA_Mat}The Cramer-Shoup cryptosystem using the semigroup
$G=M_{3\times3} \mathbb{Z}_{7}[S_5]$ is secure against adaptive
chosen ciphertext attack assuming that (1) the hash function $H$ is
chosen from a  universal one-way family, and (2) the decision
Diffie-Hellman problem is hard in the group $G$.
\end{thm}

Before beginning the proof of the theorem we need the following two
experimental facts.
\begin{enumerate}
\item Given an invertible matrix $M \in G=M_{3\times3} \mathbb{Z}_{7}[S_5]$ and random
integers $a,b$ and $c\in \mathbb{N}$, it is not possible to
distinguish between the distributions  generated by
$(M^a,M^b,M^{ab})$ and $(M^a,M^b,M^c)$.

\item Given an invertible matrix $M \in G=M_{3\times3} \mathbb{Z}_{7}[S_5]$ and a random integer
$a$,
it is not possible to extract information about $a$ from $M^a$ and
$M$.  In other words, the distributions generated by a random matrix
$N$ and $M^a$ are indistinguishable.
\end{enumerate}

We offer the following two experiments as evidence for the
plausibility of the above facts. For these tests we used invertible
matrices over the group ring $M_{3\times3} \mathbb{Z}_{7}[S_5]$. For
the first we chose a random invertible matrix $M$ (see section
\ref{Inv_Mat}) and random integers $a$, $b$ and $c\in\mathbb{N}$. We
choose $a$ and $b$ in the interval $[10^{22},10^{27})$ and $c$ in
the interval $[10^{44},10^{54})$ so that $ab$ and $c$ were roughly
of the same size.  For each pair of  resulting matrices $M^{ab}$ and
$M^c$ we counted the frequency of elements of $S_5$ appearing in
each entry.

\begin{figure}[!ht]
\includegraphics[width=0.55\textwidth]{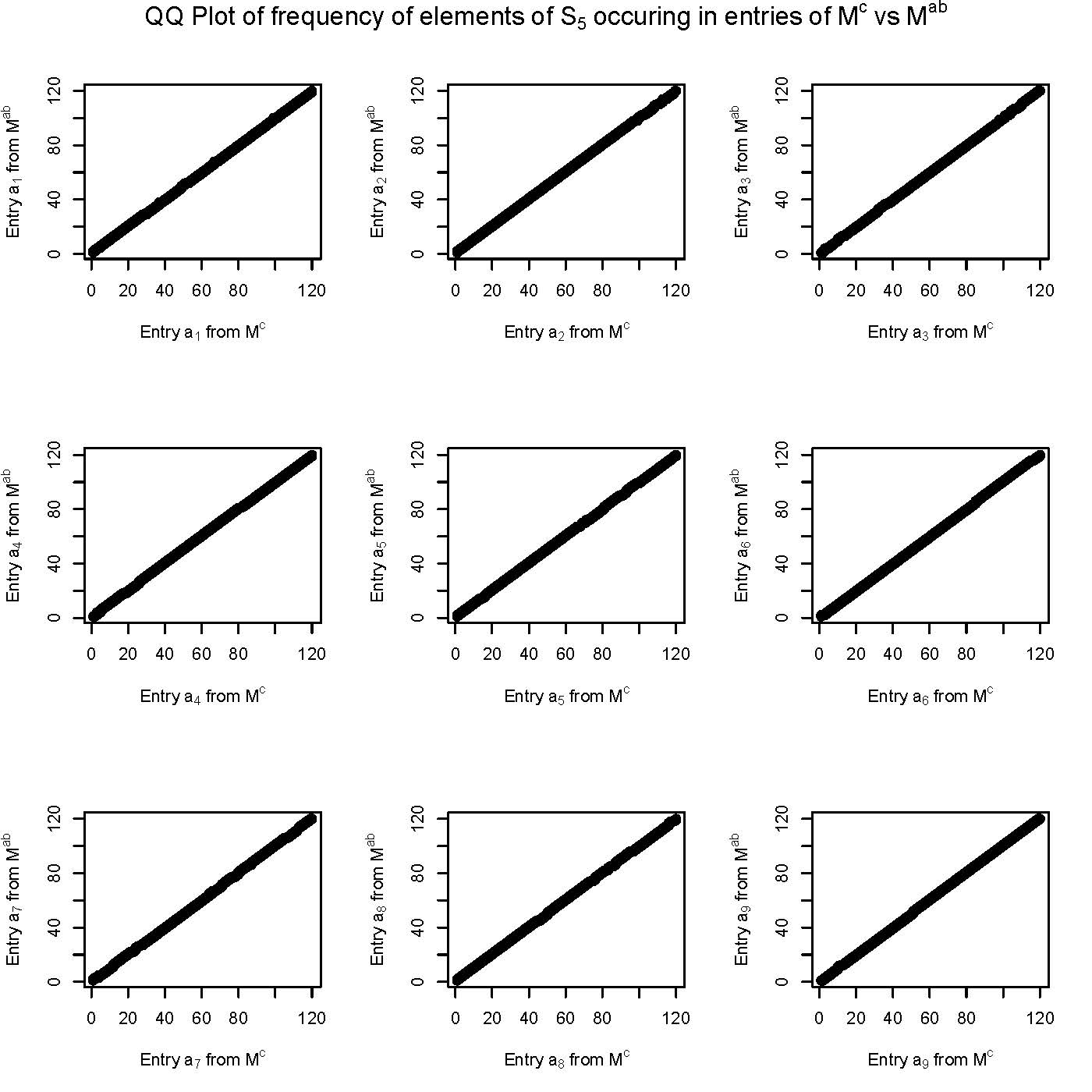}
\caption{DDH results for $M^{c}$ vs. $M^{ab}$}
\label{DDH_fig}
\end{figure}

Repeating this $500$ times for randomly chosen $a$, $b$ and $c$, we
obtained a  frequency distribution of elements of the group ring in
each entry of the two matrices. From this we created the QQ-plots
for each of the $9$ matrix entries. QQ-plots are a quick and easy
way to test for identical distributions, in which case the plots
should be straight lines. As we can see from Figure \ref{DDH_fig},
it appears that from the generated  distributions  it is not
possible to distinguish DH pairs from non-DH pairs.

\begin{figure}[!ht]
\includegraphics[width=0.55\textwidth]{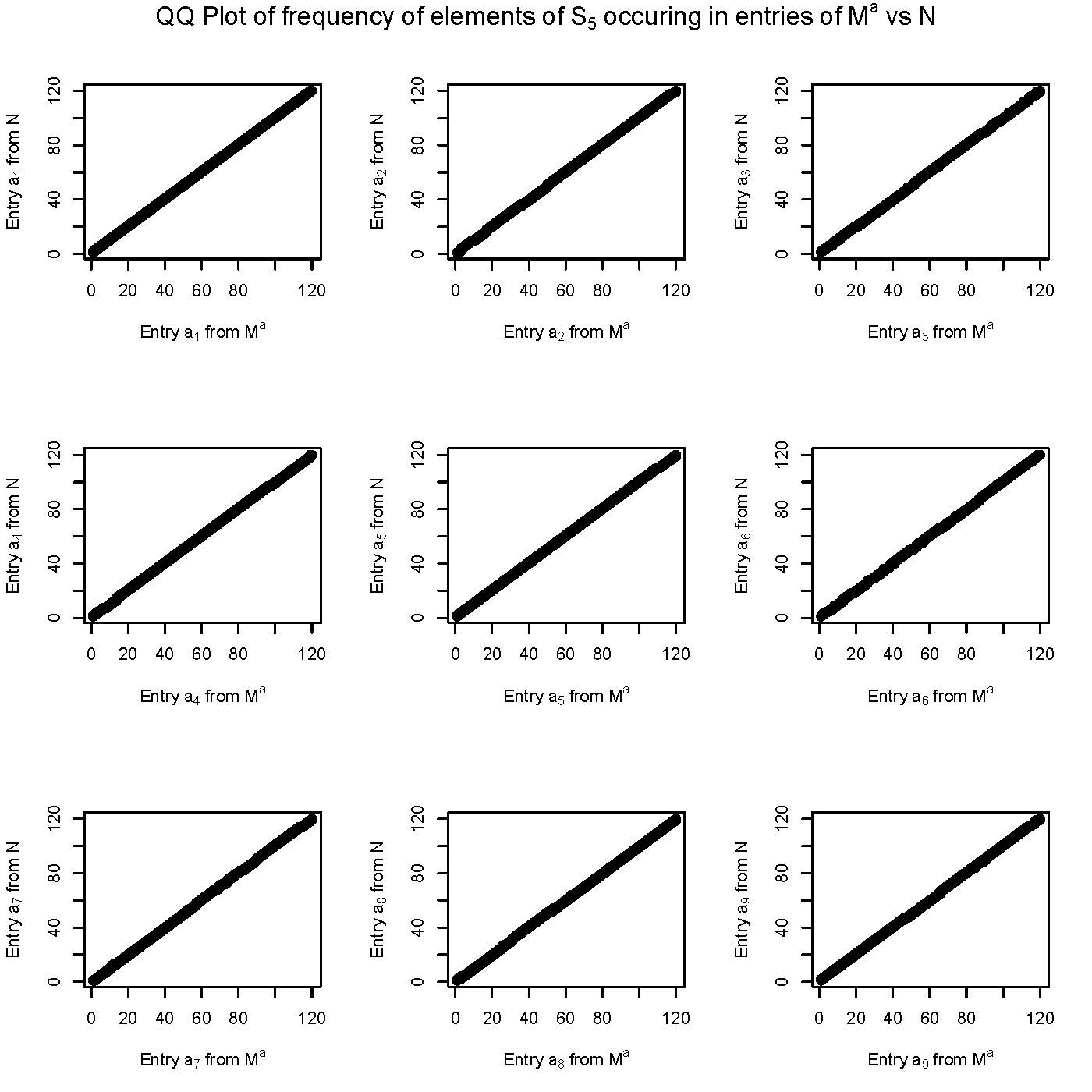}
\caption{Results for $M^a$ vs. $N$}
\label{info_plot}
\end{figure}

For verification of the second fact, we conducted a similar
experiment, except in this case, for each of the $500$ draws we
varied all parameters $N$, $M$ and $a$. We again generated QQ-plots
as shown in Figure \ref{info_plot}, and these show that no
information about $a$ is leaked from publishing $M$ and $M^a$.

We are now ready to prove Theorem \ref{CCA_Mat}. The proof will
proceed in a similar  fashion as Cramer-Shoup's original proof. We
will begin by constructing an algorithm $D$ to attack the DDH
assumption. This algorithm relies on a probabilistic polynomial time
adversary $A$ attacking our scheme, which succeeds with probability
$p$, $\mathbb{P}_A(Success)=p$. Denote by $DH$ the set of valid
Diffie-Hellman tuples $(M_1,M_2,M_1^r,M_2^r)$, and by $R$ the set of
all random tuples $(M_1,M_2,M_3,M_4)$. Then the algorithm is
constructed  as follows:
 \begin{itemize}
 \item $D$ receives input $(M_1,M_2,M_3,M_4) \text{ from } DH \text{ or } R$
 \item Pick $x_1,x_2,y_1,y_2,z \in \mathbb{Z}_n$ and a universal one-way hash function $H$
 \item The adversary $A$ receives the public key, PK, which is
 \[(M_1,M_2,c=M_1^{x_1}M_2^{x_2},d=M_1^{y_1}M_2^{y_2},h=M_1^{z},H)\]
 \item The adversary picks two messages $m_0,m_1$ and publishes them
 \item $D$ picks $b \in \{0,1\}$ and passes to $A$
 \[(M_3,M_4,M_3^z\cdot m_b,M_3^{x_1+\alpha x_2}M_4^{y_1+\alpha y_2}),\]
 where $\alpha=H(M_3,M_4,M_3^z \cdot m_b)$
 \item With this information $A$ tries to determine $b$ and returns its guess $b'$
 \item If $b=b'$ return ``DH'', else ``R''
 \end{itemize}

The proof is then verifying that this algorithm cannot attack the
$DDH$ problem. It is built from the following three claims.

\textbf{Claim 1:}
$|\mathbb{P}(D=DH|DH)-\mathbb{P}(D=DH|R)|<\epsilon$. This claim is
trivially true since $D$ is a PPT algorithm and the DDH assumption
holds as verified previously.

\textbf{Claim 2:} $\mathbb{P}(D=DH|DH)=\mathbb{P}_A(Success)$. If we
are given a DDH tuple, then all decryption queries succeed for $A$.
Hence the output of $A$ will match the choice of $b$ with
$\mathbb{P}_A(Success)$.

\textbf{Claim 3:} $|\mathbb{P}(D=DH|R)-\frac{1}{2}|<\epsilon$. Since
$\mathbb{P}(D=DH)=\mathbb{P}(A=b)$, the proof of this claim relies
on the proof of two pieces. We need to show that for all decryption
queries where $u_1=M_1^{r_1}$ and $u_2=M_2^{r_2}$ with $r_1\ne r_2$,
the decryption verification fails with  non-negligible probability.
In addition to this, we must also show that assuming all invalid
decryptions fail, the adversary A does not learn any additional
information about $z$.

We first start with the latter piece. If all invalid decryptions
fail, then the only additional information A receives is when valid
decryptions are performed. Thus, at the onset of the attack $A$ only
has information available that is given to him from PK, namely
$h=M_1^z$. If $A$ submits a valid ciphertext $(u_1',u_2',e',v')$,
where $u_1'=M_1^{r'}$, then $A$ obtains that $h^{r'}=M_1^{z^{r'}}$.
However, based on the results above, if we denote  $M=M_1^z$, then
$h^{r'}=M^{r'}$ and the distributions of any random matrix $N$ and
$M^{r'}$ generated by $r'$ are indistinguishable, hence nothing is
revealed about $z$.

Furthermore, from the encryption information passed to $A$, the only
additional information $A$ has is $M_3^z\cdot m_b$, which leaves him
with obtaining information from $M_3^z$ and $M_1^z$, i.e. solving a
Diffie-Helmann problem, which we assumed was difficult in our scheme
setup.

We are now left with showing that decryption almost always fails for
invalid ciphertexts. Suppose that the adversary submits an invalid
ciphertext, $(u_1',u_2',e',v')\ne(u_1,u_2,e,v)$. Then we have the
following cases:

\textbf{Case 1:} If $(u_1,u_2,e)=(u_1',u_2',e')$ and $v\ne v'$, then
the hash values $\alpha$ and $\alpha'$ will be the same, however
decryption will certainly be rejected.

\textbf{Case 2:} If $(u_1,u_2,e)\ne (u_1',u_2',e')$ but $a = a'$,
then this means that $A$ has found a collision in $H$. But we
assumed $H$ was collision resistant, and since $A$ runs in
polynomial time,  this can only happen with negligible probability.

\textbf{Case 3:} If $H(u_1,u_2,e)\ne H(u_1',u_2',e')$, then we have
the following system of equations where we denote by
$\log=\log_{M_1}$ and $w=\log(M_2)$, and $u_1=M_1^{r_1}$,
$u_1'=M_1^{r_1'}$, $u_2=M_2^{r_2}$ and $u_2'=M_2^{r_2'}$:
  \begin{align}
    \log c=& x_1 + wx_2 \\
    \log d =& y_1 + wy_2\\
    \log v = &r_1x_1 + wr_2x_2 + \alpha r_1y_1 + \alpha wr_2y_2 \\
    \log v' = &r_1'x_1 + wr_2'x_2 + \alpha' r_1'y_1 + \alpha' wr_2'y_2.
\end{align}

These equations are linearly independent as can be verified by looking at \\
\begin{center}
$det
\begin{pmatrix}
1 & w & 0 & 0  \\
0 & 0 & 1 & w \\
r_1 & wr_2 & \alpha r_1 & \alpha w r_2 \\
r_1' & wr_2' & \alpha' r_1' & \alpha' w r_2' \\
\end{pmatrix}
= w^2(r_2-r_1)(r_2'-r_1')(\alpha - \alpha')$
\end{center}

The above determinant is nonzero since we are considering bad
decryptions and hence
\[ r_1 \ne r_1', r_2 \ne r_2', \alpha \ne \alpha'.\]
Therefore, almost surely any bad decryption queries of this form
will be rejected.

Thus we have shown from Claim 3 that the adversary $A$ is unable to
correctly determine $b$ given a random tuple, which we saw is
equivalent to our algorithm not being able to distinguish a random
tuple from a DH tuple when given a random tuple. This together with
Claim 1 shows that our algorithm cannot distinguish between tuples
no matter what the input was. And finally, from Claim 2, we get that
the adversary is unable to attack our scheme with an adaptive chosen
ciphertext attack. \qed

\subsection{Parameters for the Cramer-Shoup-like scheme using matrices  over group rings}
Here we address  two problems relevant to key generation in our
scheme, namely, (1) how to sample  invertible matrices and (2) how
to sample commuting matrices.

\subsubsection{Invertible matrices}\label{Inv_Mat} Sampling  invertible matrices can be done
using various techniques. The first method is to construct a matrix
which is a product of  elementary matrices,
\[M = \prod_{i=1}^n E_i,\]
where $E_i$ is any elementary matrix from
$M_{3\times3}(\mathbb{Z}_7[S_5])$. Elementary  matrices can be of
one of the three types below. In the matrix $T_i(u)$, the element
$u$ should be invertible in $\mathbb{Z}_7[S_5]$.


\begin{scriptsize}
 \begin{align*}
 T_{i,j} &=
  \begin{pmatrix}
 1 & & & & & & \\
    & \ddots & & & & & \\
    &            & 0 & & 1 & & \\
    &        &    & \ddots & & & \\
    &        & 1 & & 0 & & \\
    & & & & & \ddots & \\
   & & & & & & 1 \\
  \end{pmatrix}
 T_i(u) &=
  \begin{pmatrix}
 1 & & & & & & \\
    & \ddots & & & & & \\
    &            & 1 &  & & & \\
    &        &    & u & & & \\
    &        &  & & 1 & & \\
    & & & & & \ddots & \\
   & & & & & & 1 \\
  \end{pmatrix}
 T_{i,j}(v) &=
  \begin{pmatrix}
 1 & & & & & & \\
    & \ddots & & & & & \\
    &            & 1 &  & & & \\
    &        &    & \ddots & & & \\
    &        &  v & & 1 & & \\
    & & & & & \ddots & \\
   & & & & & & 1 \\
  \end{pmatrix}
 \end{align*}
 \end{scriptsize}

We can then easily compute $M^{-1}$ as
\[M^{-1} = \prod_{i=1}^n E^{-1}_{n-i+1}\]
The drawback of generating an invertible matrix this way is that we
do not have a good grasp of the randomness embedded in this process.
In particular, how large must $n$ be to generate a truly random
matrix? Given that there are 3 different types of elementary
matrices, does it matter in what order they are multiplied in and
does the number of elementary  matrices of each form matter? These
are questions that have not been addressed and may influence the
final invertible matrix generated in unknown ways.

Here, instead of the previously mentioned method of sampling random
matrices, we propose an alternative solution. We start with an
already ``somewhat random" matrix, for which it is easy to compute
the inverse. An example of such a matrix is a lower/upper triangular
matrix, with invertible elements on the diagonal:
\begin{align*}
M&=\begin{pmatrix}
u_1 & g_1 & g_2 \\
0 & u_2 & g_3 \\
0 & 0 & u_3 \\
\end{pmatrix}.
\end{align*}

Constructing the inverse of this matrix involves solving a matrix equation,
\begin{align*}
M\cdot  M^{-1} & = I \\
\Rightarrow
\begin{pmatrix}
u_1 & g_1 & g_2 \\
0 & u_2 & g_3 \\
0 & 0 & u_3 \\
\end{pmatrix}
\cdot
\begin{pmatrix}
u_1^{-1} & g_4 & g_5 \\
0 & u_2^{-1} & g_6 \\
0 & 0 & u_3^{-1} \\
\end{pmatrix} &=
\begin{pmatrix}
1 & 0 & 0 \\
0 & 1 & 0 \\
0 & 0 & 1 \\
\end{pmatrix}\\
\Rightarrow
g_4 & = -u_1^{-1}g_1u_2^{-1} \\
g_5 & = u_{1}^{-1}g_1u_2^{-1}g_3u_3^{-1}-u_1^{-1}g_2u_3^{-1}\\
g_6 & = -u_2^{-1}g_3u_3^{-1}.\\
\end{align*}

We then propose to take a random product of such invertible upper
and lower triangular matrices. Since these matrices are more complex
than elementary matrices, it seems reasonable to assume that we
arrive at a more uniform  distribution sooner than by simply using
elementary matrices. In our experiments we used a product of 20
random matrices, where each term of the product was chosen randomly
as either a random invertible upper or lower triangular matrix.

As mentioned previously, the benefits of this method are that
inverses are  easy to compute and that the chosen matrix already has
a large degree of randomness built in. In particular, any element of
$\mathbb{Z}_7[S_5]$ can be used off the diagonal, and any invertible
elements of the group ring can be used on the diagonal. These of
course include elements such as $nu \in \mathbb{Z}_7[S_5]$, where $u
\in S_5$ and $n\in \mathbb{Z}_7$.

Finally, we note that the order of the group $GL_3\mathbb{Z}_7[S_5]$
of invertible  $3\times3$ matrices over  $\mathbb{Z}_7[S_5]$ is at
least $10^{313}$. Indeed, if we only count  invertible upper and
lower triangular matrices that we described above, then we already
have $(7\cdot 120)^3(7^{120})^3\sim 10^{313}$ matrices.

\subsubsection{Commuting matrices}

Now that we have sampled an invertible matrix ($M_1$ in our notation
-- see Section \ref{protocol}), we have to sample an arbitrary
(i.e., not necessarily invertible) matrix $M_2$ that would commute
with $M_1$.


Given a matrix $M_1\in G$, define $M_2=\sum_{i=1}^k a_iM_1^i$, where
$a_i\in \mathbb{Z}_7$ are selected randomly. Then clearly
$M_1M_2=M_2M_1$. A reasonable choice for $k$ is about 100 as this
would yield $7^{100} \sim 10^{85}$ choices for $M_2$, which is a
sufficiently large key space.

\subsubsection{Other parameters}

As mentioned in the introduction of the Cramer-Shoup algorithm
adapted to our  group rings, we need to specify the value of $n$ for
$\mathbb{Z}_n$.  Based on experiments in our previous paper
\cite{KKS} we suggest $n \sim 10^{100}$. This seemed a reasonable
choice of exponent since it both allowed quick computations and
ensured that the power a matrix was raised to could not be figured
out by brute force methods alone.

We also use a hash function $H$ in our algorithm as did Cramer and
Shoup.  The only requirement on $H$ is that it is drawn from a
family of  universal one-way hash functions. This is a less
stringent requirement than to be {\it  collision resistant}. The
latter implies that  it is infeasible for an adversary to find two
different inputs $x$ and $y$ such that $H(x)=H(y)$.  A weaker notion
of {\it second preimage resistance} implies that upon choosing an
input $x$, it is infeasible to find a different input $y$ such that
$H(x)=H(y)$.

It should be noted that in their paper Cramer and Shoup also give
details of their  same algorithm without requiring the use of any
hash functions. The modified algorithm is only slightly more
complicated but relies on the same principles.

\end{document}